\pdfoutput=1
\documentclass[10pt, conference]{IEEEtran}
\IEEEoverridecommandlockouts
\usepackage{cite}
\usepackage{amsmath,amssymb,amsfonts}
\usepackage{graphicx}
\usepackage{textcomp}
\usepackage{amsmath}
\usepackage{booktabs} 
\usepackage{multirow}
\usepackage{hyperref}       %
\usepackage{url}            %
\usepackage{booktabs}       %
\usepackage{amsfonts}       %
\usepackage{nicefrac}       %
\usepackage{microtype}      %
\usepackage{xcolor}         %
\usepackage{algorithm}
\usepackage{algpseudocode}
\usepackage{amsmath}
\usepackage{tikz}
\usetikzlibrary{arrows.meta, positioning, fit, backgrounds, calc, shapes.geometric}

\usepackage{orcidlink}
\usepackage{comment}
\usepackage{amssymb}
\usepackage{braket}
\usepackage{xcolor}
\usepackage{setspace}
\usepackage[skip=3pt, font=small]{caption}
\def\BibTeX{{\rm B\kern-.05em{\sc i\kern-.025em b}\kern-.08em
    T\kern-.1667em\lower.7ex\hbox{E}\kern-.125emX}}

\begin{document}
\bstctlcite{IEEEexample:BSTcontrol}

\title{Q-MALDetect: Quantum Image-Based Malware Detection and Adversarial Robustness Analysis}

\author{\IEEEauthorblockN{Walid El Maouaki\IEEEauthorrefmark{2}, Mohammad Izen\IEEEauthorrefmark{2}, Nouhaila Innan\IEEEauthorrefmark{2}\IEEEauthorrefmark{4}, Alberto Marchisio\IEEEauthorrefmark{2}\IEEEauthorrefmark{4}\IEEEauthorrefmark{3}$^*$\thanks{$^*$This work has been carried out while at NYUAD.},
Muhammad Shafique\IEEEauthorrefmark{2}\IEEEauthorrefmark{4}}

\IEEEauthorblockA{\IEEEauthorrefmark{2} \normalsize eBrain Lab, Division of Engineering, New York University Abu Dhabi, PO Box 129188, Abu Dhabi, UAE\\}
\IEEEauthorblockA{\IEEEauthorrefmark{4}Center for Quantum and Topological Systems (CQTS), NYUAD Research Institute, NYUAD, Abu Dhabi, UAE\\ }
\IEEEauthorblockA{\IEEEauthorrefmark{3} \normalsize SDU Microelectronics, Institute of Mechanical and Electrical Engineering, University of Southern Denmark, Odense, Denmark\\
Emails: \{walid.el.maouaki, mi2527, nouhaila.innan, muhammad.shafique\}@nyu.edu, chisio@sdu.dk}
}

\maketitle

\begin{abstract}
Quantum machine-learning (QML) models are increasingly considered for security applications, yet their adversarial robustness in malware analysis remains insufficiently understood. This paper presents Q-MALDetect, a hybrid convolutional and variational-quantum detector for byte-image representations of x86-64 Executable and Linkable Format (ELF) binaries, and evaluates it under feature-space and a range of Grad-CAM-guided, executable-aware evasion attacks. The model achieves 97.2\% training accuracy, but its robustness deteriorates as attack strength increases, with the strongest evaluated feature-space attack reaching a 78.7\% attack success rate (ASR). Among executable-aware attacks, Grad-CAM-guided no-operation (NOP) insertion produces the highest conditional ASR of 32.9\%. Dynamic adversarial training is the most effective evaluated defense, reducing feature-space ASR to as low as 5.0\%, executable-aware ASR to 7.7\%, and the sequential-attack ASR to 34.0\%. In contrast, convolutional-autoencoder preprocessing increases executable-aware ASR to 39.7\%, coinciding with a rise in Grad-CAM attribution mass within the mutable \texttt{.text} region from 35.2\% to 45.7\%. Ablation results reveal no monotonic relationship between qubit count and robustness, while the encoding interface modifies the learned attack surface without neutralizing the strongest attack. These findings demonstrate that the effectiveness of attacks and defenses depends strongly on the perturbation constraints and evaluation conditions. Reliable deployment of QML malware detectors therefore requires assessment against both feature-space and executable-aware attacks rather than reliance on clean accuracy or generic denoising alone.
\end{abstract}


\section{Introduction}
Linux malware has become a critical security concern because servers, embedded platforms, and IoT devices commonly rely on Executable and Linkable Format (ELF) instances. Static malware detectors based on deep learning can process binaries quickly and can avoid the cost of dynamic sandbox execution. A widely adopted approach converts executable bytes into images and applies convolutional neural networks (CNNs), allowing models to learn discriminative byte-layout patterns without hand-crafted features~\cite{nataraj2011malware}. However, CNNs are vulnerable to adversarial examples, i.e., small, carefully chosen perturbations can cause high-confidence misclassification~\cite{goodfellow2015explaining,madry2018towards}. In malware detection, this threat is especially subtle because arbitrary feature-space perturbations may not correspond to a valid executable. Recent ELF-focused work, such as ADVeRL-ELF, has shifted toward problem-space attacks that maintain both executability and malicious behavior while evading detection~\cite{ravi2025adverl}.

Quantum machine learning (QML) offers a new design space for security analytics because parameterized quantum circuits can act as trainable feature maps with representations that differ from classical neural networks. Prior studies have investigated the robustness of quanvolutional and hybrid quantum-classical models~\cite{advqunn2024,robqunns2024,robustqnnmetrics2025}. Yet most QML robustness studies use canonical image data, while adversarial malware research focuses predominantly on classical models. Consequently, it remains unclear \textit{how a hybrid QML malware detector behaves when exposed to both unconstrained feature-space perturbations and executable-aware binary mutations.}

This paper addresses that question through a vulnerability study of \textit{Q-MALDetect}, a hybrid convolutional and variational-quantum ELF classifier built on established hybrid QML detector designs~\cite{barrue2023quantum,liu2023can,wang2025network}; the contribution is a joint feature-space and executable-aware robustness evaluation of the complete pipeline. This study asks four questions: (i) whether the hybrid model fits the clean data to a high-recall operating point; (ii) which feature-space and executable-aware attacks are effective, and under what conditions; (iii) whether adversarial training and CAE preprocessing provide consistent protection; and (iv) how perturbation budget, qubit count, encoding interface, and simulated hardware noise affect the observed attack surface.

The main contributions are:
\begin{itemize}
\item We formulate a critical-domain robustness evaluation for a hybrid quantum-classical ELF malware detector and separate clean classification, feature-space sensitivity, and executable-aware evasion.
\item We benchmark a spectrum of attacks that combines FGSM, PGD, and an FGSM-initialized PGD variant with two Grad-CAM-guided NOP-insertion attacks. The latter attacks use attribution to align a constrained mutation budget with model-sensitive executable regions; one selects NOP actions randomly and one tests whether tabular Q-learning improves on it.
\item We compare two standard defenses, dynamic adversarial training and CAE preprocessing, and identify an asymmetric defense outcome: the CAE attenuates additive feature-space perturbations but can increase Grad-CAM-guided problem-space ASR by shifting attribution toward the mutable \texttt{.text} region.
\item We evaluate adversarial training and ablations over perturbation budget, qubit count, and encoding strategy, showing that quantum-interface choices materially affect attack success.
\end{itemize}

\section{Background and Related Work}

Byte-image malware classification maps executable bytes into grayscale or color images and applies image classifiers~\cite{nataraj2011malware}. Classical malware evasion studies have moved from unrestricted image perturbations toward functionality-aware binary transformations. COPYCAT injects adversarial content into malware representations~\cite{khormali2019copycat}; reinforcement-learning approaches search over executable transformations~\cite{anderson2018learning}; EvilELF studies structural modifications to Linux binaries~\cite{kosikowski2023evilelf}; and ADVeRL-ELF inserts semantic NOP sequences into ELF code regions to evade an image-based ResNet detector~\cite{ravi2025adverl}. These studies demonstrate that executable constraints usually reduce attack freedom but produce a more operationally relevant threat model.

Quantum methods have been studied for malware, intrusion, network-security, and side-channel analysis tasks~\cite{barrue2023quantum,liu2023can,abreu2024qml,wang2025network,quantumclustering2024,bhatnagar2026meta,elmaouaki2026qleak,maouaki2026qgcl}. In parallel, adversarial QML research has established attacks, training-based defenses, robustness verification, universal perturbations, and experimental demonstrations on quantum hardware~\cite{liao2021robust,du2021quantum,guan2021robustness,weber2021optimal,gong2022universal,ren2022experimental,innan2025quantum}. More recent studies investigate architecture, expressibility, transferability, circuit metrics, and federated robustness~\cite{advqunn2024,robqunns2024,robustqnnmetrics2025,qfal2025,robqfl2025,elmaouaki2026qrail}. These results motivate attack-aware QML design, but they do not establish robustness for executable malware.

\textbf{Research gap.} Existing work does not provide a unified evaluation of a hybrid QML malware detector under both differentiable feature-space attacks and attribution-guided executable mutations, nor does it examine whether defenses can protect against one threat class while exposing another. This paper focuses on that gap.

\section{Study Design and Methodology}

\begin{figure*}[t]
    \centering
    \includegraphics[width=\linewidth]{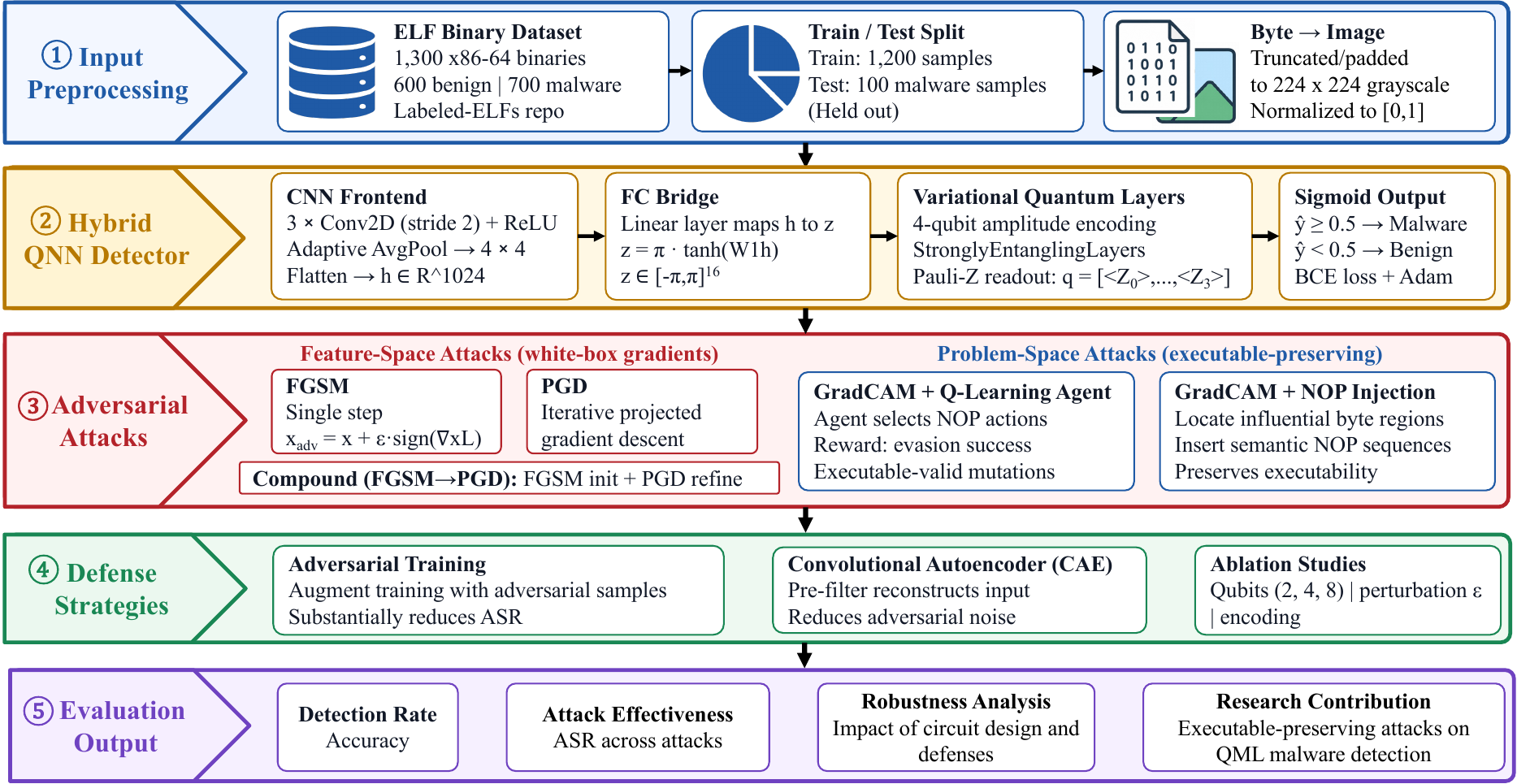}
    \caption{Q-MALDetect evaluation workflow. ELF binaries are converted to $224\times224$ normalized byte images and divided into a 1,200-sample model-development pool and a 100-malware attack holdout. A CNN bridge supplies normalized amplitudes to a variational quantum circuit. The evaluation covers three feature-space attacks and two Grad-CAM-guided, executable-aware NOP-insertion attacks; dynamic adversarial training and CAE preprocessing are then tested under the same attack families. Ablations vary perturbation budget, qubit count, encoding interface, and NOP length, while the final stage reports clean metrics, ASR, attribution behavior, and noise-model sensitivity.}
    \label{method}
\vspace{-3mm}
\end{figure*}

Figure~\ref{method} summarizes the complete protocol, from ELF preprocessing and hybrid inference through attacks, defenses, ablations, and evaluation. The protocol maintains separate model-development and held-out attack pools and distinguishes feature-space perturbations from reconstructed binary mutations.

\subsection{Threat Model and Dataset}
We study non-targeted evasion: an adversary starts from malware that the model classifies correctly and seeks a benign prediction. All attacks assume access to the trained model. FGSM and PGD require end-to-end gradients. The Grad-CAM-guided attacks require gradients for attribution and model-confidence feedback for evaluating candidate mutations, restricted to semantics-neutral NOP insertion with ELF repair. 

The dataset contains 1,300 x86-64 ELF binaries from the Labeled-ELFs repository~\cite{labeledelfs}, including 600 benign and 700 malware samples. The training set contains all 600 benign samples and 600 malware samples. The remaining 100 malware samples are held out from training and reserved exclusively for adversarial attack evaluation. Clean metrics are computed on the balanced training set and fix each model's operating point; every ASR is measured on these unseen malware, matching the malware-to-benign evasion threat model.

\begin{figure}[t]
  \centering
  \includegraphics[width=\columnwidth]{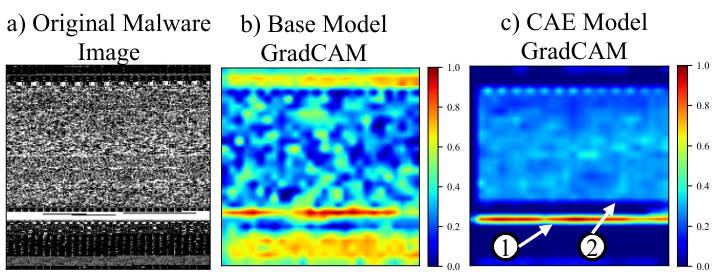}
  \caption{Representative attribution shift under CAE preprocessing. (a) Original malware byte image. (b) The base-model Grad-CAM map is spatially diffuse. (c) The CAE-model map concentrates high activation in a narrow horizontal band aligned with the analyzed \texttt{.text} region. The two numbered callouts mark segments of this same high-activation band, not separate processing stages. Across 495 paired maps, \texttt{.text} attribution mass increases from $35.2\%$ to $45.7\%$.}
  \label{fig:gradcamcae}
\vspace{-3mm}
\end{figure}

Each file is represented by its first $224\times224=50{,}176$ bytes. Short files are zero-padded and long files are truncated; byte values are divided by 255 to obtain $\mathbf{x}\in[0,1]^{224\times224}$. Truncation means that attacks and explanations concern only file regions visible within the first 50,176 bytes. Figure~\ref{fig:gradcamcae} illustrates both the resulting byte image in Figure~\ref{fig:gradcamcae}(a) and a Grad-CAM map in Figure~\ref{fig:gradcamcae}(b) used to guide domain-space attacks.

Clean performance is reported using accuracy, precision, malware recall, and false-positive rate. Attack success rate is defined over initially correct malware:
\begin{equation}
\mathrm{ASR}=\frac{\#\{i:f(\mathbf{x}_i)=1,\ f(\mathbf{x}^{\mathrm{adv}}_i)=0\}}
{\#\{i:f(\mathbf{x}_i)=1\}}\times100\%,
\label{eq:asr}
\end{equation}
where label 1 denotes malware. Restricting the denominator to initially correct samples prevents clean-model errors from being counted as attack successes. Results are summarized as mean $\pm$ standard deviation across seeds.

\subsection{Hybrid Quantum-Classical Model Detector}

The CNN frontend contains three stride-2 convolutional layers with ReLU activations, followed by adaptive average pooling to $4\times4$ and flattening to $\mathbf{h}\in\mathbb{R}^{1024}$. For the default four-qubit amplitude-encoding model, the fully connected bridge produces
\begin{equation}
\mathbf{z}=\pi\tanh(\mathbf{W}_1\mathbf{h}),
\qquad \mathbf{z}\in[-\pi,\pi]^{16}.
\end{equation}
Amplitude embedding requires a unit-norm vector, so the circuit input is
\begin{equation}
\widetilde{\mathbf{z}}=
\frac{\mathbf{z}}{\max(\lVert\mathbf{z}\rVert_2,\delta)},
\end{equation}
with a small numerical constant $\delta>0$. A single PennyLane \texttt{StronglyEntanglingLayers} block~\cite{bergholm2018pennylane} processes the four-qubit state. Four Pauli-$Z$ expectation values are passed to a linear layer and sigmoid to produce malware probability $\hat{y}$; the decision threshold is 0.5. The CNN and bridge use learning rate 0.001, while the quantum parameters and output head use 0.01. Training uses Adam, binary cross-entropy, batch size 32, and 10 epochs.

\subsection{Feature-Space Attacks}

FGSM applies one loss-maximizing step and clips the result to the valid input range:
\begin{equation}
\mathbf{x}_{\mathrm{FGSM}}=
\Pi_{[0,1]}\!\left(\mathbf{x}+\varepsilon\,
\mathrm{sign}\!\left(\nabla_{\mathbf{x}}\mathcal{L}(\mathbf{x},y)\right)\right),
\end{equation}
with $\varepsilon=0.01$ in the main experiment.

PGD applies 50 steps with $\alpha=0.001$ and projects after every update:
\begin{equation}
\mathbf{x}^{(t+1)}=
\Pi_{\mathcal{B}_{\infty}(\mathbf{x},\varepsilon)\cap[0,1]}
\!\left(\mathbf{x}^{(t)}+\alpha\,
\mathrm{sign}\!\left(\nabla_{\mathbf{x}^{(t)}}\mathcal{L}\right)\right).
\label{eq:pgd}
\end{equation}

The compound attack applies FGSM first and then runs PGD from the FGSM output. In the reported implementation, the second projection is centered on that intermediate point. Each stage uses the same nominal $\varepsilon$, so the final displacement from the original input can reach $2\varepsilon$ before clipping. The compound result is therefore a stronger sequential threat model, not a same-budget comparison with FGSM or PGD. These attacks modify normalized byte-image values and are not asserted to correspond to valid executable files.

\subsection{Executable-Aware Grad-CAM Attacks}
\subsubsection{Single-Point Insertion and Reconstruction}
The two domain attacks use Grad-CAM~\cite{selvaraju2017gradcam} at the third convolutional layer to rank influential image locations. Candidate locations are mapped back to byte offsets, restricted to the ELF \texttt{.text} section, and snapped to a nearby instruction boundary identified through Capstone disassembly. Boundaries crossed by short relative branches are rejected. All NOPs for one sample accumulate at a single accepted boundary, reducing the repair problem to one cut position and one total shift.

After inserting a block of $N$ bytes, the procedure expands \texttt{.text}, shifts subsequent section and program-header offsets and virtual addresses, updates relevant ELF header fields, and adjusts relative branches, calls, RIP-relative references, relocations, symbols, and dynamic-table addresses that cross the cut. LIEF is then used to reparse the rebuilt binary. Since multi-byte NOPs leave the architectural state unchanged and all listed cross-cut references are repaired, the mutation is semantics-neutral by design; every rebuild is verified for structural validity and re-read consistency, and smoke tests on rebuilt benign utilities confirm runnability. An \emph{executable-aware} success is thus a structurally valid, semantics-neutral rebuild classified as benign.

Let $\mathcal{I}$ denote initially correct attack malware and $\mathcal{R}\subseteq\mathcal{I}$ the samples for which a safe boundary is found and a structurally valid rebuild is produced. The plotted domain metric is
\begin{equation}
\mathrm{ASR}_{\mathrm{exec}}=
\frac{\left|\{i\in\mathcal{R}:f(\mathbf{x}^{\mathrm{rebuilt}}_i)=0\}\right|}
{|\mathcal{R}|}\times100\%.
\label{eq:asr_exec}
\end{equation}

The total success rate is $\text{ASR}_{\text{exec}} \times |R|/|I|$. The executable rate ($\text{ASR}_{\text{exec}}$) acts as the maximum possible overall rate, achieved at 100\% reconstruction coverage.
Feature-space and executable-aware ASRs measure distinct vulnerabilities and are not directly comparable.

\subsubsection{Grad-CAM-Guided Random NOP Insertion}
The random baseline selects among six architecturally valid multi-byte x86-64 NOP encodings of length 6-11 bytes. At each turn, it inserts one randomly selected NOP at the accumulating Grad-CAM-guided boundary, re-reads the first 50,176 bytes, and queries the detector. The attack stops after a label flip or 200 turns. A successful image-level evasion is counted only after reconstruction, reparsing, and rescoring of the rebuilt ELF.

\subsubsection{Grad-CAM-Guided Q-Learning}
The learned variant tests whether selecting a particular NOP encoding improves on random choice.
The state is one of 10 confidence buckets,
\begin{equation}
 s=\min\left(9,\left\lfloor10c\right\rfloor\right),
\end{equation}
where $c\in[0,1]$ is the current malware confidence. The six NOP encodings form the action space. With
\begin{equation}
 p=\mathrm{clip}\!\left(
 \frac{c_{\mathrm{orig}}-c_{\mathrm{new}}}
 {c_{\mathrm{orig}}-0.5+\delta},-1,1\right),
\end{equation}
the reward is
\begin{equation}
 r=\begin{cases}
 10, & c_{\mathrm{new}}<0.5,\\
 2p, & p>0,\\
 p, & p\leq0.
 \end{cases}
\end{equation}
The Q-table update is
\begin{equation}
Q(s,a)\leftarrow Q(s,a)+\alpha_Q
\left[r+\gamma\max_{a'}Q(s',a')-Q(s,a)\right],
\end{equation}
with $\alpha_Q=0.1$ and $\gamma=0.9$. Exploration uses $\epsilon_{\mathrm{RL}}$-greedy selection, decaying from 0.9 to 0.05 by a factor of 0.995 per episode. Policy fitting uses eight passes over model-development malware, with at most 200 turns per episode. The frozen policy acts greedily on the attack holdout, and only reconstructed evasions contribute to Eq.~\eqref{eq:asr_exec}.

\begin{table}[t]
\centering
\caption{Experimental configuration summary.}
\label{tab:exp_setup}
\small
\begin{tabular}{ll}
\toprule
\textbf{Category} & \textbf{Configuration} \\
\midrule
Dataset & 1300 ELF ($224\times224$) \\
Split & 600 benign, 700 malware \\
Training & 10 epochs, batch 32, Adam, BCE \\
Learning rate & CNN 0.001, quantum 0.01 \\
Architecture & CNN + FC + QNN + FC \\
Default circuit & 4 qubits, 1 layer, Amplitude \\
Gradient attacks & FGSM, PGD, Compound \\
FGSM & $\varepsilon=0.01$ \\
PGD & $\varepsilon=0.01$, $\alpha=0.001$, 50 iter \\
Domain attacks & GradCAM+NOP, GradCAM+RL \\
Insertion & single safe \texttt{.text} boundary \\
NOP set & 6 canonical NOPs, 6--11 bytes \\
Turns per sample & up to 200 \\
Reconstruction & Capstone + LIEF, structure-valid \\
Reported ASR & image (upper bound) + executable \\
RL agent & tabular Q-learning, 8 passes \\
RL state / action & 10 conf.\ buckets / 6 NOPs \\
Adv.\ training & dynamic PGD/epoch + NOP/RL aug., 5 ep. \\
CAE & 4-layer, latent 128, 20 epochs \\
Qubit ablation & $n \in \{2,3,4,5,6,7,8,9\}$ \\
Encoding ablation & Amplitude, Angle \\
Epsilon ablation & $\varepsilon \in [10^{-3},10^{1}]$, 10 pts \\
Seeds & 3 seeds for all experiments unless %
specified\\
\bottomrule
\end{tabular}
\vspace{-1mm}
\end{table}

\subsection{Defenses, Ablations, and Noise-Model Evaluation}

To improve Q-MALDetect robustness, we apply adversarial training (AT) following the min-max formulation of Madry \emph{et al.}~\cite{madry2018towards}. At each of five fine-tuning epochs, fresh PGD examples are generated against the current model and replace 50\% of the corresponding development samples. Static reconstructed NOP and RL bypass examples generated from model-development malware are also included so that hardening covers both threat classes. The held-out attack malware remain excluded from fine-tuning.

As a second line of defense, we prepend a Convolutional Autoencoder (CAE) to the Q-MALDetect pipeline as a preprocessing stage. The CAE consists of a four-layer convolutional encoder that progressively downsamples the $224\times224$ image to a $14\times14\times128$ representation, which is compressed into a 128-dimensional latent vector. A symmetric four-layer transposed-convolution decoder reconstructs the input with a sigmoid output. The CAE is trained for 20 epochs on clean model-development data using mean-squared error and Adam with a learning rate of 0.001. It is then frozen, and the classifier is trained on reconstructed inputs. All white-box feature-space attacks differentiate through the complete CAE-classifier pipeline.

We also study the impact of three design variables on adversarial vulnerability. The qubit sweep uses $n\in\{2,3,4,5,6,7,8,9\}$. Because amplitude encoding requires $2^n$ inputs and the readout returns $n$ expectation values, this sweep grows the bridge from 4,096 to 524,288 weights ($1024\cdot2^n$) while the circuit grows from 6 to 27 rotation angles ($3n$). The encoding study compares amplitude embedding with four-feature angle embedding at four qubits; it therefore evaluates complete encoding interfaces, in which the bridge dimension (16 vs.\ 4), gradient flow, and encoding gates change jointly. The perturbation study evaluates 10 logarithmically spaced $\varepsilon$ values from $10^{-3}$ to $10^1$. Since inputs are clipped to $[0,1]$, budgets at or above 1 are nominal values that primarily expose saturation behavior rather than progressively larger feasible $\ell_\infty$ regions.

A final transfer experiment evaluates adversarial inputs generated against the noiseless model under 1,024-shot FakeManilaV2 simulation, a local backend whose noise model is derived from a device snapshot; it thus measures sensitivity to stored device noise and finite-shot sampling. %
Table~\ref{tab:exp_setup} consolidates the default settings across the main comparison and ablations.

All classical components are implemented in PyTorch. PennyLane provides the differentiable quantum simulation, while LIEF and Capstone support ELF parsing and disassembly.

\section{Results and Discussion}
\label{sec:results}

\subsection{Clean Performance and Undefended Attack Surface}

\begin{figure}[t]
\centering
\includegraphics[width=\columnwidth]{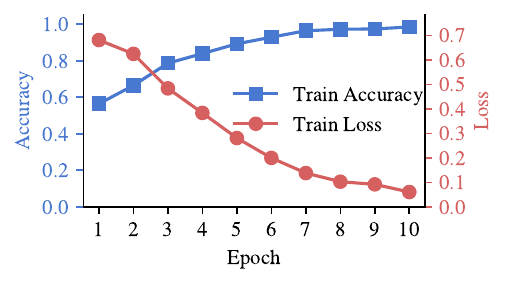}
\caption{Training convergence for one illustrative run. Loss decreases from 0.6815 to 0.0605 while training accuracy increases from $56.41\%$ to $98.44\%$ over 10 epochs. %
}
\label{fig:convergence}
\vspace{-3mm}
\end{figure}

The illustrative run in Fig.~\ref{fig:convergence} shows stable optimization: the training loss decreases from 0.6815 at epoch 1 to 0.0605 at epoch 10, while the training accuracy increases from $56.41\%$ to $98.44\%$. Across the evaluated runs, the model achieves mean training-set accuracy of $97.2\%\pm1.0\%$, precision of $95.1\%\pm1.4\%$, malware recall of $99.6\%\pm0.8\%$, and a false-positive rate of $5.2\%\pm1.5\%$. These results confirm that the model fits the clean training data and provides a high-recall starting point for the robustness analysis.

\begin{figure}[t]
  \centering
  \includegraphics[width=\columnwidth]{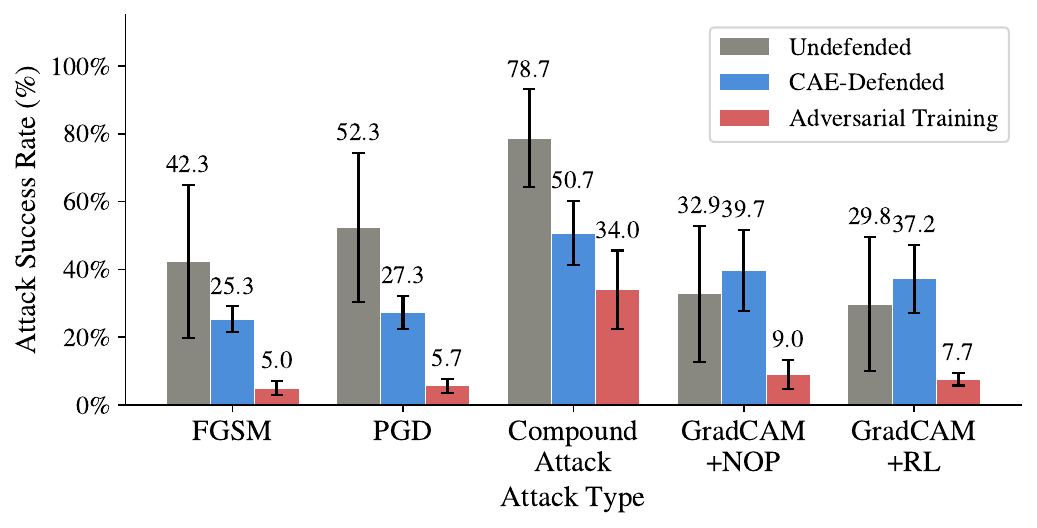}
  \caption{ASR under the undefended, CAE-defended, and adversarially trained conditions. FGSM, PGD, and Compound are feature-space ASRs; Grad-CAM+NOP and Grad-CAM+RL are conditional executable ASRs. Bar labels round the corresponding means to whole percentages. AT reduces all five attacks, whereas the CAE reduces feature-space ASR but increases both executable-aware rates.}
  \label{fig:threeway}
\vspace{-3mm}
\end{figure}

In Figure~\ref{fig:threeway}, under the undefended model, FGSM reaches $42.3\%$ ASR and PGD reaches $52.3\%$. The sequential compound attack is strongest at $78.7\%$, consistent with its two-stage optimization and potentially doubled displacement budget. 

The executable-aware attacks attain conditional ASRs of $32.9\%$ for Grad-CAM+NOP and $29.8\%$ for Grad-CAM+RL. These rates are lower than the compound feature-space result but operationally more concrete: each counted success is a reparsed, semantics-neutral rebuild that the detector scores as benign. Comparisons with prior ELF attacks~\cite{kosikowski2023evilelf,xue2024elf,ravi2025adverl,hrdonka2026elf} must account for different models, architectures, mutation sets, budgets, denominators, and functionality checks.

The random and learned NOP attacks perform similarly: the 3.1-point mean difference in favor of random selection lies within the seed variation, although the Q-tables learn state-dependent preferences (mean within-state action spread of 2.2 value units, 1.6--3.0 across seeds). These preferences cannot raise ASR because the actions change NOP form and length but not the insertion location or the dominant image-layout effect.

\begin{figure}[t]
  \centering
  \includegraphics[width=\columnwidth]{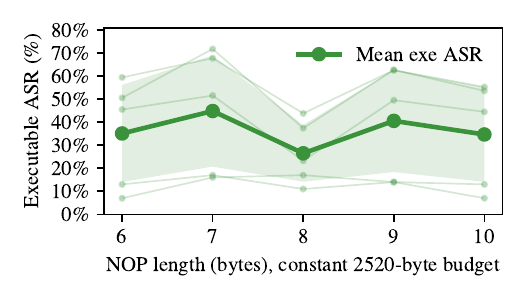}
  \caption{Conditional executable ASR versus NOP length with total inserted volume fixed at 2,520 bytes, averaged over five seeds. Faint lines show individual seeds and the bold line shows the mean. The absence of a monotonic length trend, together with large seed variation, indicates that total displacement and detector fragility dominate the identity of the selected NOP.}
  \label{fig:nopsize}
\vspace{-3mm}
\end{figure}

The fixed-volume diagnostic in Figure~\ref{fig:nopsize}, an ablation over constant-action policies at matched volume, supports this interpretation. With total inserted volume held at 2,520 bytes, mean conditional executable ASR is $35.1\%$, $44.8\%$, $26.5\%$, $40.5\%$, and $34.6\%$ for NOP lengths 6-10. The sequence is non-monotonic and the between-seed spread is larger than any consistent length trend. Inserting bytes at one early \texttt{.text} boundary shifts all subsequent visible bytes, so total layout displacement and the chosen boundary provide a stronger mechanism than the semantic identity of a particular NOP. An effective learned attacker should therefore control location, volume, or transformation family.

\subsection{Defense Effectiveness}
\subsubsection{Dynamic Adversarial Training}
\begin{figure*}[t]
\centering
\includegraphics[width=\textwidth]{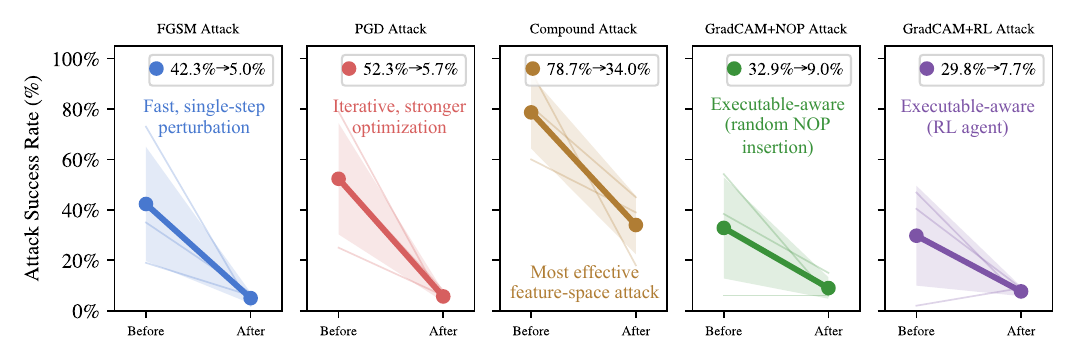}
\caption{Paired ASR before and after dynamic adversarial training. Faint lines show individual seeds, bold lines show means, and shaded regions show $\pm1$ standard deviation. AT sharply reduces both feature-space and executable-aware attacks, although the sequential compound attack retains a $34.0\%$ mean ASR.}
\label{fig:advtraining}
\vspace{-3mm}
\end{figure*}

Dynamic AT is the strongest defense in Figure~\ref{fig:threeway}. FGSM drops from $42.3\%$ to $5.0\%$ (a 37.3-point reduction), PGD from $52.3\%$ to $5.7\%$ (46.6 points), and the compound attack from $78.7\%$ to $34.0\%$ (44.7 points). Grad-CAM+NOP falls from $32.9\%$ to $9.0\%$, and Grad-CAM+RL from $29.8\%$ to $7.7\%$. Figure~\ref{fig:advtraining} shows that these reductions occur across the paired seeds rather than being produced by one outlying run.

The result follows the composition of the fine-tuning data: fresh PGD examples harden the differentiable attack surface, while reconstructed NOP/RL examples expose the model to domain mutations. The residual $34.0\%$ compound ASR remains important. PGD-based hardening does not fully cover a sequential attack whose feasible displacement can exceed the single-stage training budget. These gains quantify resistance to malware evasion, the objective of the threat model; the clean-accuracy and false-positive cost of hardening is the complementary deployment metric.

\subsubsection{Convolutional-Autoencoder Preprocessing}
The CAE produces a threat-dependent reversal. Relative to the undefended model, FGSM falls from $42.3\%$ to $25.3\%$, PGD from $52.3\%$ to $27.3\%$, and the compound attack from $78.7\%$ to $50.7\%$. The bottleneck removes part of the additive image-space perturbation, but it does not eliminate the sequential attack.

For executable-aware attacks, the same preprocessing is harmful: Grad-CAM+NOP rises from $32.9\%$ to $39.7\%$, and Grad-CAM+RL rises from $29.8\%$ to $37.2\%$. The attribution analysis provides a coherent mechanism. Across 495 paired maps, the fraction of Grad-CAM activation mass inside \texttt{.text} increases from $35.2\%$ to $45.7\%$, a 10.5-point mean shift, and 60\% of samples move in that direction. As Figure~\ref{fig:gradcamcae} illustrates, the CAE model exhibits pronounced activation confined to a narrow band that is spatially aligned with the mutable code region (in panel c), as indicated by circular markers 1 and 2 in the figure. This does not by itself prove causation, but it is consistent with the measured increase in attacks that deliberately target \texttt{.text}. The observation complements interpretation-based ELF defenses~\cite{qiao2023interpretation}: attribution can identify adversarial structure, but a defense that changes attribution can also expose a more concentrated mutation target.

\subsection{Sensitivity and Ablation Studies}
\subsubsection{Perturbation Budget}
\begin{figure}[t]
\centering
\includegraphics[width=\columnwidth,trim=0 5 0 5,clip]{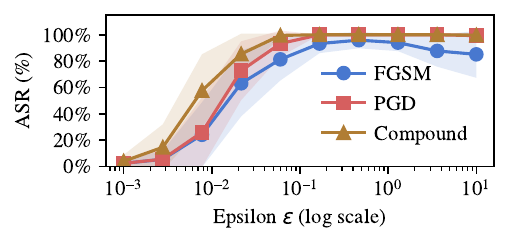}
\caption{Feature-space ASR versus nominal perturbation budget for FGSM, PGD, and the sequential compound attack. %
Shaded bands denote $\pm1$ standard deviation. The 10 values are logarithmically spaced from $10^{-3}$ to $10^1$; values at or above 1 primarily reveal clipping-induced saturation.}
\label{fig:epsilon}
\vspace{-3mm}
\end{figure}

Figure~\ref{fig:epsilon} evaluates a ten-seed run over 10 logarithmically spaced budgets. In the transition between approximately $\varepsilon=$ $8\times10^{-3}$ and $2\times10^{-2}$, ASR rises sharply and uncertainty is widest; the compound attack increases first, followed by PGD and FGSM. PGD and the compound attack approach saturation only over the subsequent points, roughly $6\times10^{-2}$ to $2\times10^{-1}$, rather than at $\varepsilon=0.02$ exactly. FGSM plateaus in the mid-80\% range and declines slightly at the largest nominal budgets.

The high-budget regime must be read with input clipping. For $\varepsilon\geq1$, the $[0,1]$ cube, rather than the nominal $\ell_\infty$ radius, determines the feasible input. The curves therefore describe sign saturation and iterative clipping, not meaningful growth to a radius of 10 in normalized byte space. The slight FGSM decline is not evidence that very large perturbations restore robustness. 

\subsubsection{Qubit Count}
\begin{figure}[t]
\centering
\includegraphics[width=\columnwidth]{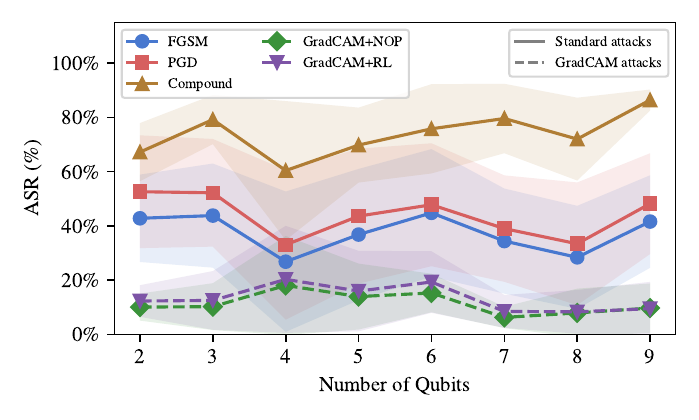}
\caption{ASR across 2-9 qubits. %
Solid curves denote feature-space attacks and dashed curves denote conditional executable ASR. No attack exhibits a monotonic qubit trend; the compound attack remains highest at every width. Because the bridge and readout dimensions change with $n$, the plot is an architecture-width ablation rather than an isolated test of quantum resources.}
\label{fig:qubits}
\vspace{-3mm}
\end{figure}

Figure~\ref{fig:qubits} shows no monotonic relation between circuit width and adversarial robustness. FGSM and PGD fluctuate substantially across 2-9 qubits, with local lows around the four- and eight-qubit configurations and renewed increases at nine qubits. The compound attack remains the largest curve at every width, ranging approximately from 60\% to 87\%. The domain attacks remain lower in this cohort, mostly within 5-20\%, but also vary non-monotonically.

The sweep therefore characterizes architecture width rather than quantum resources in isolation: changing $n$ simultaneously changes the amplitude-input dimension from $2^n$, the number of measured expectations, parameter count, and optimization geometry. Wide uncertainty bands further show that initialization and data partitioning are consequential. This architecture-coupled behavior is consistent with prior QML robustness studies in which circuit design and learned geometry, rather than qubit count alone, govern attackability~\cite{advqunn2024,west2023benchmarking,robustqnnmetrics2025}. 

\subsubsection{Encoding Interface}
\begin{figure}[t]
  \centering
  \includegraphics[width=\columnwidth]{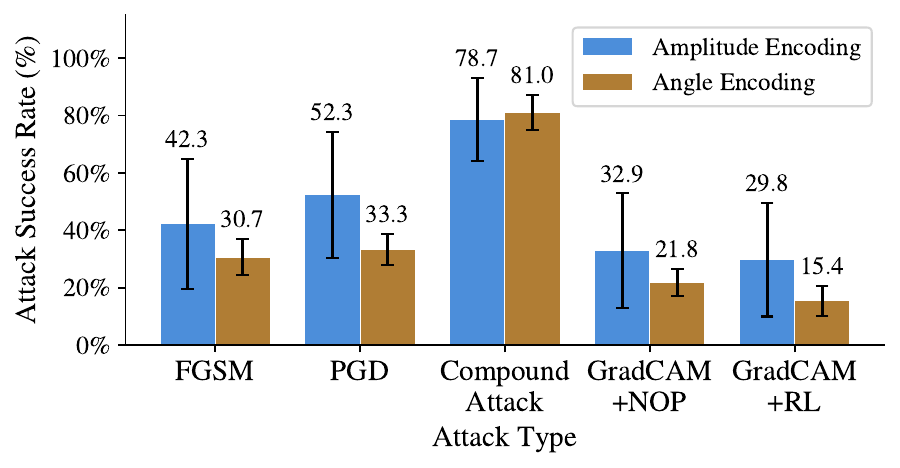}
  \caption{ASR for amplitude- and angle-encoded four-qubit models. %
  The angle-encoded interface lowers FGSM, PGD, and both conditional executable rates, while the sequential compound attack remains approximately 80\%. The bridge dimension changes with the encoding, so the comparison covers the complete interface rather than encoding gates alone.}
  \label{fig:encoding}
\vspace{-3mm}
\end{figure}

At four qubits, amplitude and angle encoding both attain $99.0\%$ training accuracy, %
but they induce different attack surfaces (Figure~\ref{fig:encoding}). The angle-encoded interface lowers FGSM from $42.3\%$ to $30.7\%$ and PGD from $52.3\%$ to $33.3\%$. It does not protect against the compound attack, which changes from $78.7\%$ to $81.0\%$. Conditional executable ASR also falls, from $32.9\%$ to $21.8\%$ for NOP insertion and from $29.8\%$ to $15.4\%$ for RL.

The domain changes are model-dependent rather than independent of encoding: changing the bridge and circuit changes the classifier gradients and thus the Grad-CAM insertion site. Because amplitude encoding receives 16 bridge features and angle encoding four, feature dimensionality, gradient flow, and the optimization landscape change together with the gates; the result therefore characterizes the complete angle-encoded interface, which is less vulnerable to four of the five attacks. The compound result demonstrates that encoding choice alone is insufficient.

\subsubsection{Shot-Based Noise-Model Sensitivity}
\begin{figure}[t]
\centering
\includegraphics[width=\columnwidth]{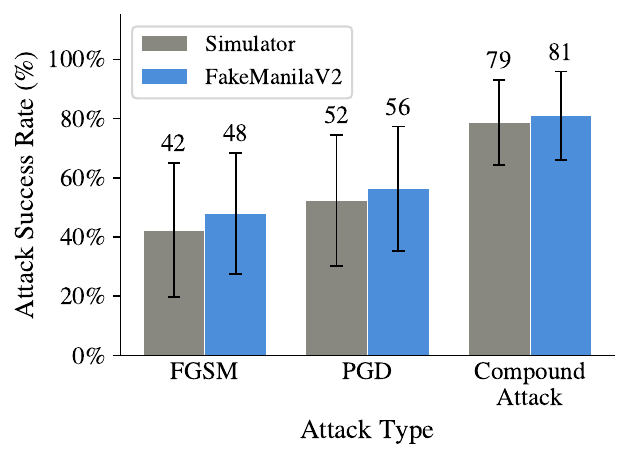}
\caption{Feature-space ASR under noiseless simulation and 1,024-shot FakeManilaV2 noise-model simulation. %
Mean shifts are at most 6.0 percentage points, while the attack ranking is unchanged. FakeManilaV2 is a local simulator backend, not physical-hardware execution.}
\label{fig:noise}
\vspace{-3mm}
\end{figure}

Figure~\ref{fig:noise} compares noiseless inference with FakeManilaV2 shot-based simulation. %
FGSM changes from $42.0\%$ to $48.0\%$, PGD from $52.0\%$ to $56.0\%$, and the compound attack from $79.0\%$ to $81.0\%$. The largest mean shift is 6 points. The ranking is preserved: compound remains strongest, followed by PGD and FGSM.

The transferred adversarial examples therefore remain effective under device-derived noise and finite sampling. Prior work shows that noise can alter quantum-classifier robustness under specific assumptions~\cite{du2021quantum}; here, no uniform protective or harmful effect appears, and the mean attack ranking is stable.

\subsection{Interpretation}
Three conclusions emerge from the combined results. First, threat-model validity changes the meaning of ASR. Feature-space attacks reveal differentiable sensitivity, but they do not produce executable malware. The Grad-CAM attacks operate under substantially tighter binary constraints, yet their conditional executable rates remain nontrivial. Recent ELF work similarly shows that effective evasion depends on where and how a binary can be transformed~\cite{xue2024elf,ravi2025adverl,hrdonka2026elf}. In Q-MALDetect, the fixed-volume experiment indicates that byte displacement at an influential boundary matters more than which NOP encoding is selected, explaining why the six-action RL policy does not outperform random choice.

Second, defense quality is threat-dependent. Dynamic AT performs best because its training distribution explicitly includes both gradient and reconstructed domain examples. The CAE is mismatched: it filters part of the additive perturbation but concentrates the classifier's evidence in the same mutable section targeted by the executable-aware attacks. Robust preprocessing must therefore be evaluated against adaptive, problem-space mutations, not only against the perturbations it was designed to remove.

Third, the quantum design variables do not yield a simple robustness law. Neither more qubits nor a different encoding guarantees protection, and the sequential attack remains strong across configurations. This agrees with the broader adversarial-QML literature: robustness is a property of the full data-encoding-circuit-optimizer system~\cite{liao2021robust,west2023benchmarking,advqunn2024,robustqnnmetrics2025}.

\section{Conclusion}
Q-MALDetect demonstrates that a hybrid quantum-classical malware detector with high training accuracy can retain a broad adversarial attack surface. The undefended model reaches $78.7\%$ ASR under the stronger sequential feature-space attack and conditional executable ASRs of $32.9\%$ and $29.8\%$ under Grad-CAM-guided NOP variants. Dynamic adversarial training reduces all five attacks, although the compound attack remains at $34.0\%$. CAE preprocessing lowers feature-space ASR but raises executable-aware ASR while shifting attribution toward \texttt{.text}, establishing a concrete defense mismatch. Qubit count has no monotonic robustness effect, angle encoding does not neutralize the compound attack, and FakeManilaV2 simulation preserves the attack ranking without constituting hardware validation. Robust QML malware detection therefore requires an explicit, multi-level threat model, reconstruction and functionality accounting, defense-specific adaptive testing, and matched classical comparisons; clean accuracy or generic denoising alone cannot support a security claim.

\section*{Acknowledgment}
 This work was supported in part by the NYUAD Center for Quantum and Topological Systems (CQTS), funded by Tamkeen under the NYUAD Research Institute grant CG008, and the Center for Cyber Security (CCS), funded by Tamkeen under the NYUAD Research Institute Award G1104.
\scriptsize{
\bibliographystyle{IEEEtran}
{\setstretch{0.88} \scriptsize{
\bibliography{refs}}}

\end{document}